\documentclass[useAMS,usenatbib]{mn2e}
\usepackage{epsfig}
\usepackage{amsmath}
\usepackage{enumerate}

\title[Probing the accretion disk structure by the ...]
{Probing the accretion disk structure by the twin kHz QPOs and spins  of neutron stars  in LMXBs}

\author[D. H. Wang et al.]
{D. H. Wang$^{1,2}\thanks{wangdh@gznu.edu.cn, zhangcm@bao.ac.cn}$, C. M. Zhang$^3$, Y. J. Lei$^3$, L. Chen$^4$, J. L. Qu$^5$, Q. J. Zhi$^{1,2}$
\\
$^1$School of Physics and Electronic Science, Guizhou Normal University, Guiyang, 550001, China\\
$^2$NAOC-GZNU Astronomy Research and Education Center, Guizhou Normal University, Guiyang, 550001, China\\
$^3$National Astronomical Observatories, Chinese Academy of Sciences, Beijing, 100012, China\\
$^4$Astronomy Department, Beijing Normal University, Beijing, 100875, China\\
$^5$Institute of High Energy Physics, Chinese Academy of Sciences, Beijing, 100049, China\\
}
\date{Released 201608}

\pagerange{\pageref{firstpage}--\pageref{lastpage}} \pubyear{2016}

\begin{document}

\maketitle

\label{firstpage}

\begin{abstract}

We analyze the relation between the emission radii of twin kilohertz quasi-periodic oscillations (kHz QPOs) and the co-rotation radii of the 12 neutron star low mass X-ray binaries (NS-LMXBs) which are simultaneously detected with the twin kHz QPOs and NS spins. We find that  the average co-rotation radius of these sources  is $\langle r_{\rm co}\rangle\sim32$\,km, and all the emission positions of twin kHz QPOs lie inside the co-rotation radii, indicating that the twin kHz QPOs are formed in the spin-up process.  It is noticed that the upper  frequency of twin kHz QPOs  is higher  than NS spin frequency by $\geq10$\%, which may
account for a critical  velocity difference between the Keplerian motion of accretion matter and  NS spin that is corresponding
to the production of twin kHz QPOs. In addition, we also find that $\sim83$\% of twin kHz QPOs  cluster around the radius range of $15-20$\,km, which may be affected by the hard surface or the local strong magnetic field of NS.
  As a special case,  SAX J1808.4-3658 shows the larger emission radii of twin kHz QPOs of $r\sim21-24$\,km, which may be due to its low accretion rate or small  measured NS mass ($<1.4\,{\rm M_\odot}$).
  
\end{abstract}


\section{Introduction}

Kilohertz quasi-periodic oscillations (kHz QPOs) are the particular phenomena in neutron star low mass X-ray binaries
 (NS-LMXBs) \citep{van der Klis06,Liu07,Walter15} and were firstly discovered by Rossi X-ray Timing Explorer (RXTE) \citep{van der Klis96,Strohmayer96}. These high-frequency QPOs usually occur in pairs (i.e.  upper $\nu_2$ and lower $\nu_1$) with the frequency range of $\simeq100-1200$\,Hz (see \citealt{van der Klis00,van der Klis06,van der Klis16} for a review), and have been detected in all subclasses of NS-LMXB, i.e. the less luminous Atoll and high luminous  Z sources (see \citealt{Hasinger89} for the Atoll and Z definitions). Such a fast X-ray variability is a powerful tool to explore the effects of general relativity in a strong gravity regime \citep{Miller98,Stella99a,Miller15},
 constrain the NS Mass-Radius relation \citep{Miller98,Miller02,Zhang04,Zhang13} and probe the accreting flow and magnetosphere-disk structure in LMXBs \citep{Kluzniak90,Kluzniak01,Abramowicz03a,Abramowicz03b,Alpar12,Peille14}.

The frequencies of the twin kHz QPOs show a nonlinear relation \citep{Belloni05,Zhang06a,Belloni07},  and
 the properties of them are also correlated with other timing and spectral features,
such as the positions in the X-ray color-color diagram (e.g., \citealt{Wijnands97a,Wijnands97b,Homan02}),
the photon indexes of the energy spectrum \citep{Kaaret98},
the noise features \citep{Ford98c},
the X-ray luminosity \citep{Mendez99b,Ford00}. In addition,  the quality factors and $rms$ amplitudes of the kHz QPOs
are found to be dependent of  the QPO frequency as well (e.g., \citealt{Mendez01,Wang12}).
Moreover, the lower kHz QPO  frequency correlates with the low-frequency (i.e. HBO, see \citealt{van der Klis06}) and they follow
a tight relation, which has been also found in the accreting white dwarf binaries \citep{Psaltis99,Belloni02,Warner02,Mauche02}.

Various theoretical models suggest that kHz QPOs reflect the orbital motion of matter at some preferred radius close to NS in LMXBs \citep{Miller98,Stella99a,Osherovich99,Lamb01,Zhang04}, and their frequencies are identified with various characteristic frequencies in the inner accretion flows or their resonances \citep{Kluzniak01,Abramowicz03a,Abramowicz03b,Torok05,Stuchlik15}.

The relativistic precession model \citep{Stella99a,Stella99b} and  Alfv\'en wave oscillation model \citep{Zhang04} emphasize the influence of the strong gravitational field regime and magnetic field near NS, respectively, which have made
 the consistent description of the model with the observed data \citep{Wang13a}.
 
The emission position of the kHz QPOs provides a probe into the physical environment near the NS in LMXBs. \citet{Wang15} analyze the relation between the emission radius of the kHz QPOs and the NS radius based on the Alfv\'en wave oscillation model, and find that most kHz QPOs emit at the position several kilometers away from the NS surface.
Besides these, the relation between the emission radius of   kHz QPOs and  co-rotation radius of NS is helpful to understand the relative velocity   between the accretion flow   and NS spin, which can further be used to investigate  the accretion environment of NS-LMXBs that arises the kHz QPOs. There are $\sim30$ LMXBs to have shown NS spins \citep{van der Klis16}, some of which have been detected with the spin period derivative \citep{Burderi06,Burderi13,Walter15}.  There are   a dozen of  NS-LMXBs to show  both  the twin kHz QPOs and NS spins (see Table \ref{QPOs}), from which the emission radii of kHz QPOs and   co-rotation radii of the sources can be inferred. The goal of this paper is to  investigate  the emission environments  of kHz QPOs while comparing with  the co-rotation radius of NS, and infer the production mechanisms  of twin kHz QPOs.

The structure of the paper is as follows: In $\S$ 2, we introduce the twin kHz QPOs and NS spin data adopted in analysis. In $\S$ 3 we infer the emission radii of kHz QPOs and analyze its relation with the co-rotation radius. In $\S$ 4 we present the discussions and conclusions.

\section{The sample of published twin kHz QPO frequencies and NS spin frequencies}
We searched  the published literature for the sources with both the detected twin kHz QPO frequencies and  NS spin frequencies, and
found that 12 sources satisfy the  above conditions. These  samples have been detected with 201 pairs of twin kHz QPOs£¬ as shown in Table \ref{QPOs} with the references,   where the 26 pairs are
taken from the accreting millisecond X-ray pulsars, and the 175 ones from Atoll sources.
 The NS spin frequencies  of the 12 sources are taken  from either periodic or nearly periodic X-ray burst oscillations (van der Klis 2000, 2006).

\begin{table*}
\begin{minipage}{150mm}
\caption{Emission radii of twin kHz QPOs and co-rotation radii of NS-LMXBs}
\setlength{\tabcolsep}{3pt}
\begin{tabular}{@{}lcccccccl@{}}
\hline
\noalign{\smallskip}
Source$^{[a]}$ (12) & $\nu_1$$^{[b]}$ & $\nu_2$$^{[c]}$ & \centering{$\nu_{\rm s}$$^{[d]}$} & $r^{[e]}$ & $r_{\rm co}$$^{[f]}$ & $Y^{[g]}$ & $\delta r$$^{[h]}$ & References \\
 & (Hz) & (Hz) & (Hz) & (km) & (km) & -- & (km) & \\
 & & & & & & ($\equiv\frac{r}{r_{\rm co}}$) & ($\equiv r_{\rm co}-r$) & \\
\noalign{\smallskip}
\hline
\noalign{\smallskip}
AMXP (2) \\
SAX J1808.4-3658 & $435\sim567$ & $599\sim737$ & 401~(AN) & $21\sim24$ & $31$ & $0.67\sim0.77$ & $7\sim10$ & [1, 13] \\
XTE J1807.4-294 & $106\sim370$ & $337\sim587$ & 191~(A) & $25\sim36$ & $53$ & $0.47\sim0.68$ & $17\sim28$ & [2, 13] \\
\noalign{\smallskip}
\hline
\noalign{\smallskip}
Atoll (10) \\
4U 0614+09 & $153\sim843$ & $449\sim1162$ & 415~(N) & $16\sim30$ & $31$ & $0.50\sim0.95$ & $2\sim16$ & [3, 14] \\
4U 1608-52 & $473\sim867$ & $799\sim1104$ & 619~(N) & $16\sim20$ & $24$ & $0.68\sim0.84$ & $4\sim8$ & [4, 13] \\
4U 1636-53 & $529\sim979$ & $823\sim1228$ & 581~(N) & $15\sim20$ & $25$ & $0.61\sim0.79$ & $5\sim10$ & [5, 13] \\
4U 1702-43 & 722 & $1055$ & 330~(N) & $17$ & $37$ & $0.46$ & $20$ & [6, 13] \\
4U 1728-34 & $308\sim894$ & $582\sim1183$ & 363~(N) & $16\sim25$ & $34$ & $0.45\sim0.73$ & $9\sim19$ & [7, 13] \\
4U 1915-05 & $224\sim707$ & $514\sim1055$ & 270~(N) & $17\sim27$ & $42$ & $0.40\sim0.65$ & $15\sim25$ & [8, 13] \\
Aql X-1 & $795\sim803$ & $1074\sim1083$ & 550~(AN) & $\sim17$ & $26$ & $0.64$ & $\sim9$ & [9, 13] \\
IGR J17191-2821 & $681\sim870$ & $1037\sim1185$ & 294~(N) & $16\sim17$ & $40$ & $0.39\sim0.43$ & $23\sim24$ & [10, 13] \\
KS 1731-260 & $898\sim903$ & $1159\sim1183$ & 524~(N) & $\sim16$ & $27$ & $0.58\sim0.59$ & $\sim11$ & [11, 13] \\
SAX J1750.8-2900 & $936$ & $1253$ & 601~(N) & $15$ & $25$ & $0.61$ & $10$ & [12, 13] \\
\noalign{\smallskip}
\hline
\noalign{\smallskip}
\end{tabular}
\label{QPOs}
\end{minipage}
\begin{tabular}{@{}l@{}}
\footnotesize
\begin{minipage}{150mm}
\begin{enumerate}[]
\item $^{[a]}$ Source with both the detected twin kHz QPOs and the inferred NS spin frequency.
\item $^{[b]}$ $\nu_1$--- Frequency of the lower kHz QPO.
\item $^{[c]}$ $\nu_2$--- Frequency of the upper kHz QPO.
\item $^{[d]}$ $\nu_{\rm s}$--- NS spin frequency inferred from periodic or nearly periodic X-ray
oscillations. A: accretion-powered millisecond pulsar. N: nuclear-powered millisecond pulsar.
\item $^{[e]}$ $r$--- Emission radius of the twin kHz QPOs inferred by equation (\ref{r_K}) (i.e. $r=(\frac{GM}{4\pi^2})^{1/3}\nu_2^{-2/3}$ with $M$ is assumed to be $1.6\,\rm M_\odot$).
\item $^{[f]}$ $r_{\rm co}$--- Co-rotation radius inferred by equation (\ref{r_co}) (i.e. $r_{\rm co}=(\frac{GM}{4\pi^2})^{1/3}\nu_{\rm s}^{-2/3}$ with $M$ is assumed to be 1.6\,$\rm M_\odot$).
\item $^{[g]}$ $Y$--- Position parameter ($Y\equiv\frac{r}{r_{\rm co}}$, see equation (\ref{Y})).
\item $^{[h]}$ $\delta r$--- $\delta r\equiv r_{\rm co}-r$.
\end{enumerate}
REFERENCES.---
[1] \citealt{van Straaten05},  \citealt{Wijnands03}, \citealt{Bult15};
[2] \citealt{Linares05}, \citealt{Zhang06b};
[3] \citealt{van Straaten00}, \citealt{van Straaten02}, \citealt{Boutelier09};
[4] \citealt{van Straaten03}, \citealt{Barret05a}, \citealt{Jonker00a}, \citealt{Mendez98b};
[5] \citealt{Altamirano08b}, \citealt{Wijnands97a}, \citealt{Bhattacharyya10}, \citealt{Di Salvo03}, \citealt{Jonker00a}, \citealt{Jonker02a}, \citealt{Lin11}, \citealt{Sanna14};
[6] \citealt{Markwardt99};
[7] \citealt{Di Salvo01}, \citealt{van Straaten02}, \citealt{Strohmayer96}, \citealt{Migliari03}, \citealt{Jonker00a}, \citealt{Mendez99};
[8] \citealt{Boirin00};
[9] \citealt{Barret08};
[10] \citealt{Altamirano10a};
[11] \citealt{Wijnands97};
[12] \citealt{Kaaret02};
[13] Reference in \citealt{Boutloukos08a};
[14] \citealt{Strohmayer08a}.
\end{minipage}
\end{tabular}
\label{QPO}
\end{table*}

\begin{figure*}
\centering
\includegraphics[width=5.5cm]{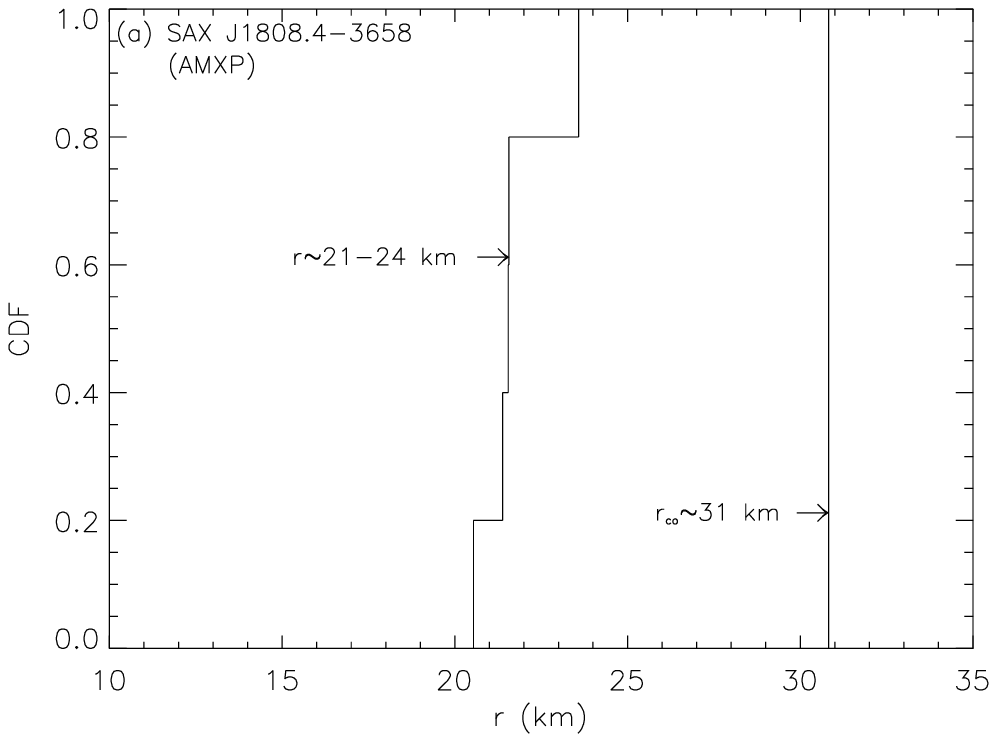}
\includegraphics[width=5.5cm]{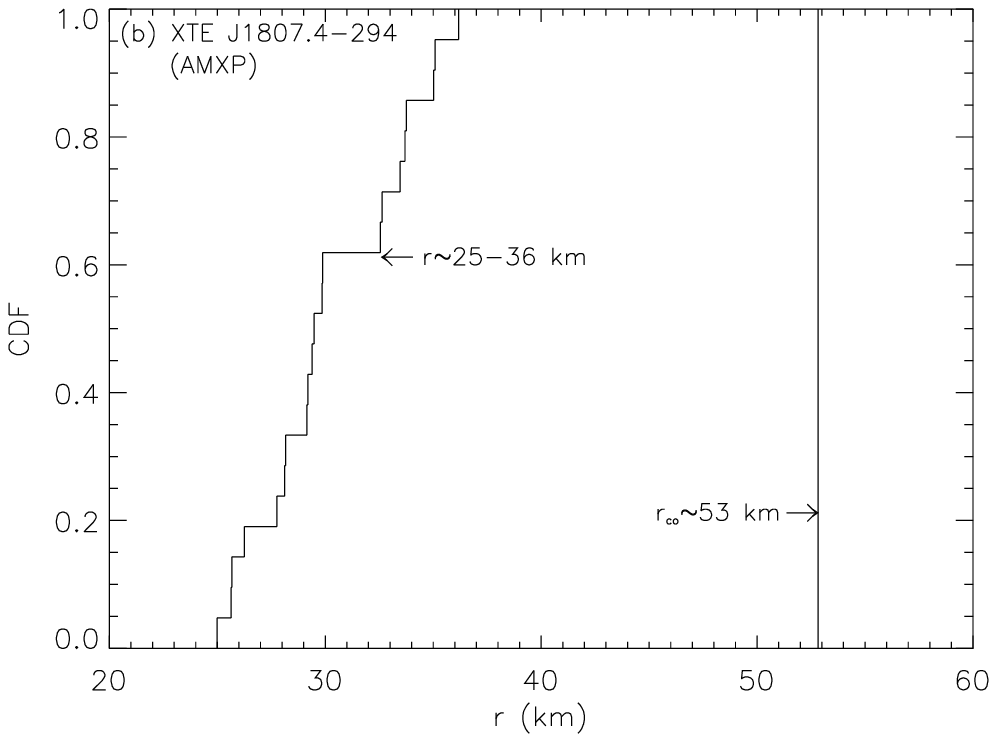}
\includegraphics[width=5.5cm]{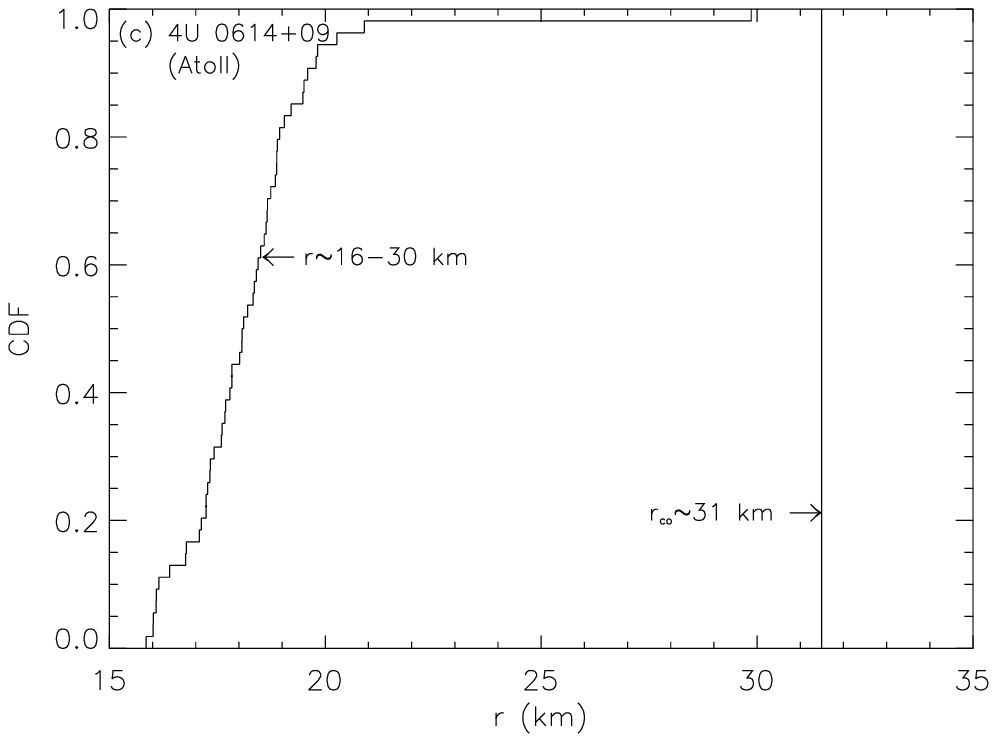}
\includegraphics[width=5.5cm]{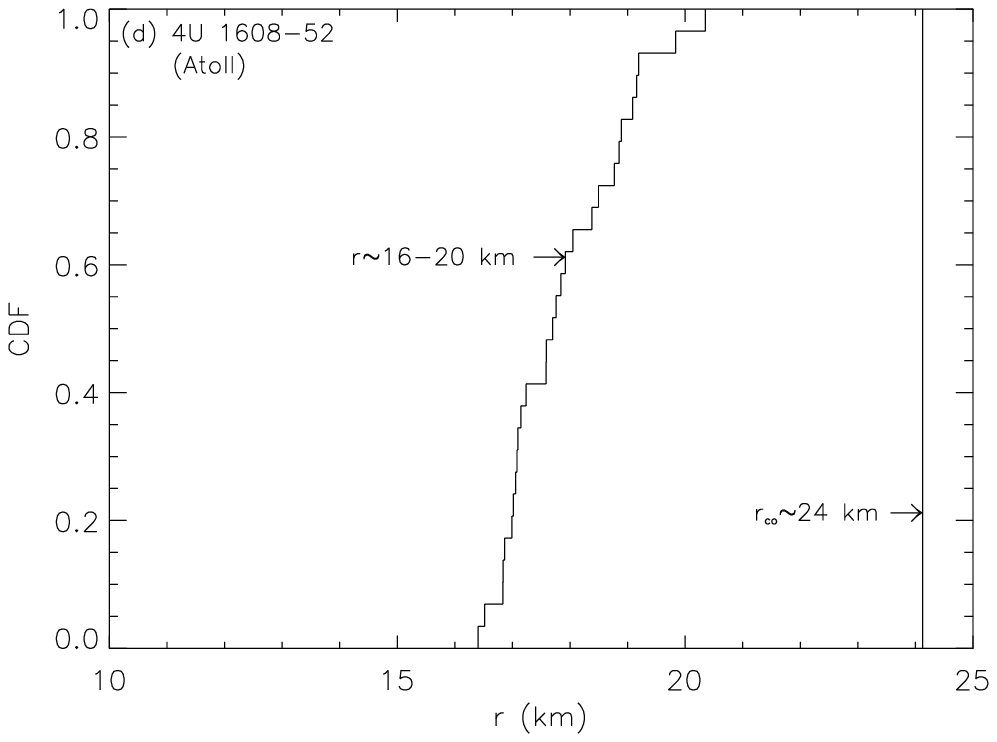}
\includegraphics[width=5.5cm]{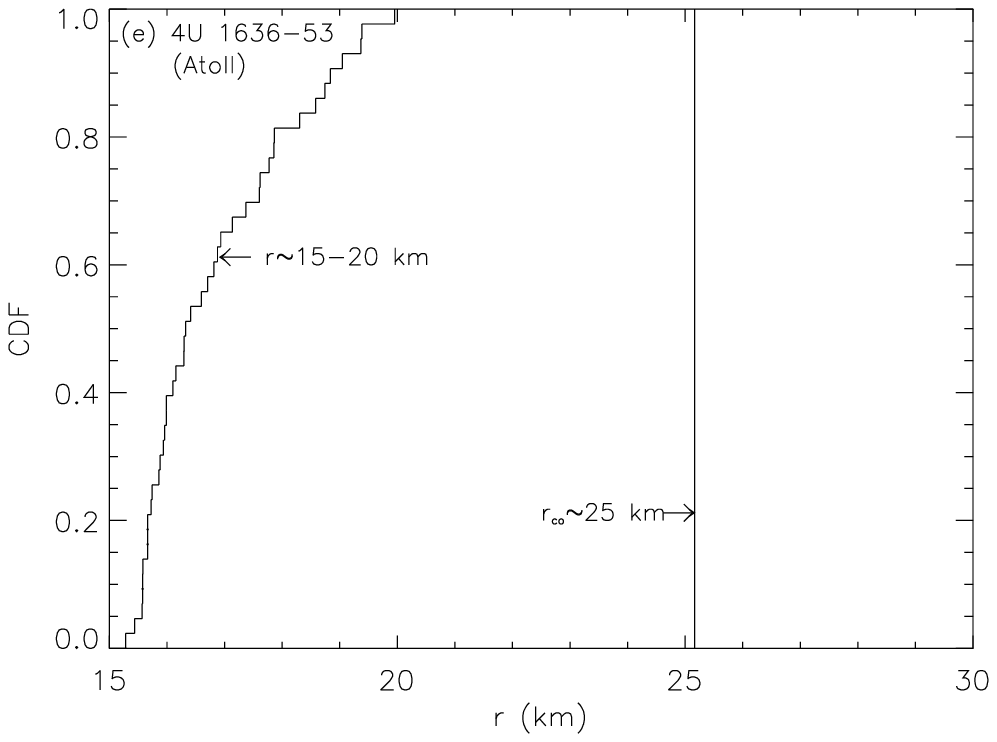}
\includegraphics[width=5.5cm]{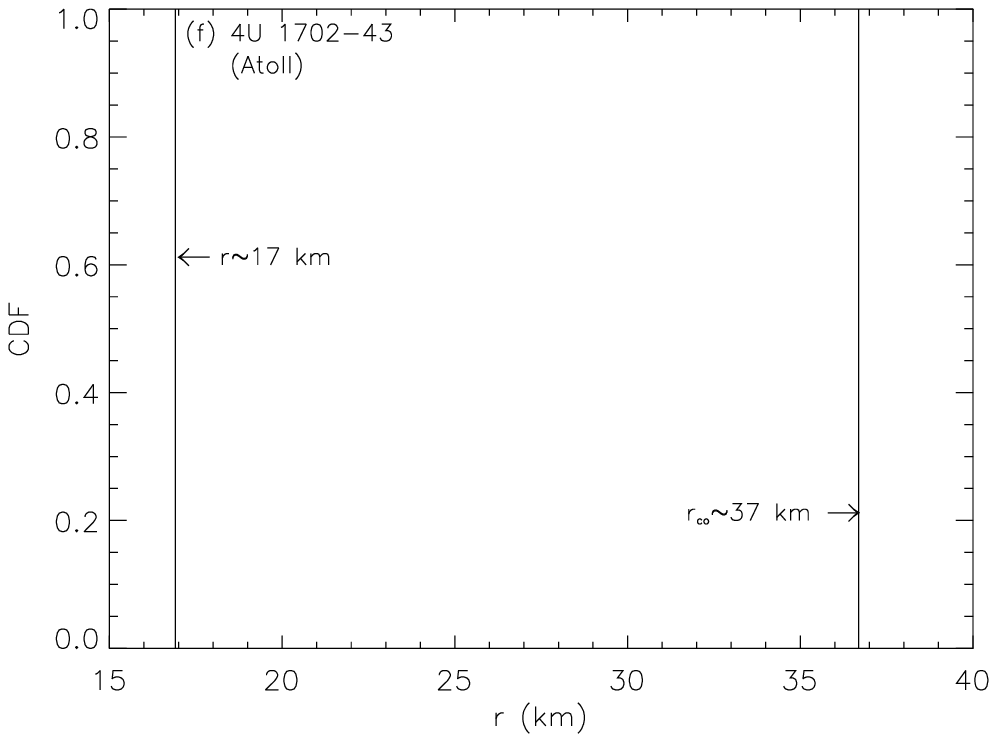}
\includegraphics[width=5.5cm]{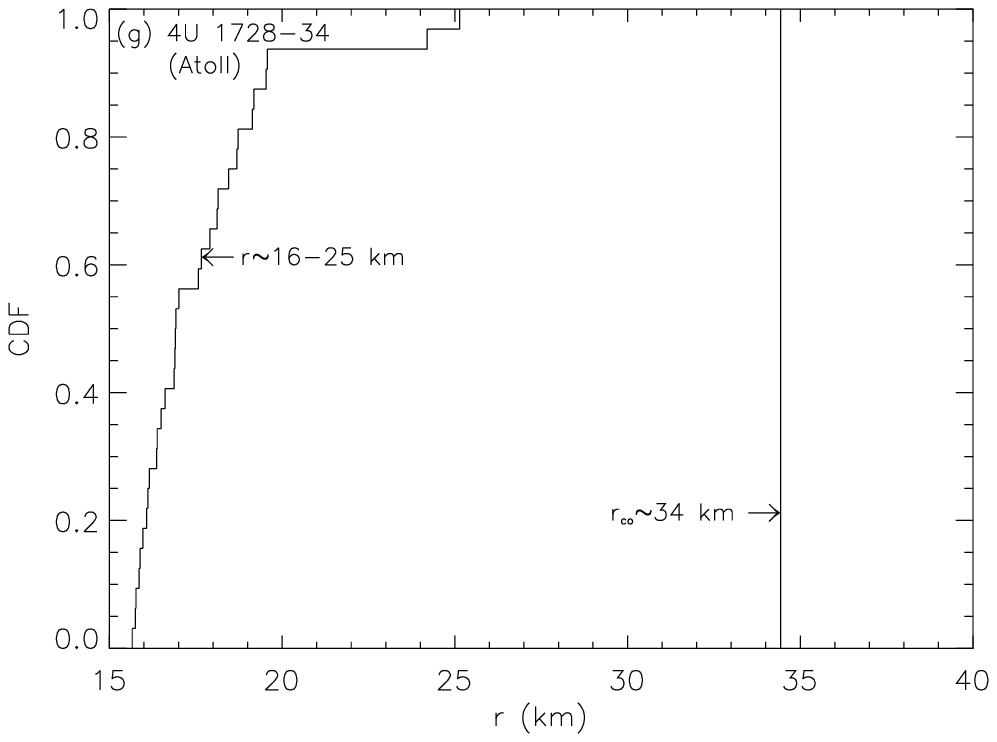}
\includegraphics[width=5.5cm]{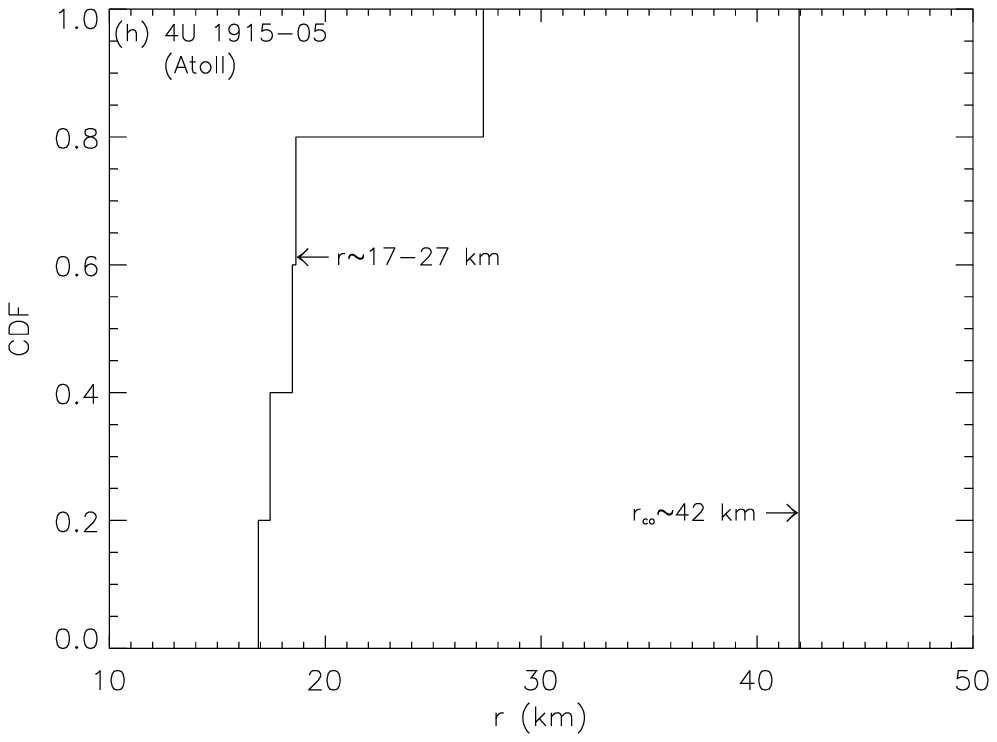}
\includegraphics[width=5.5cm]{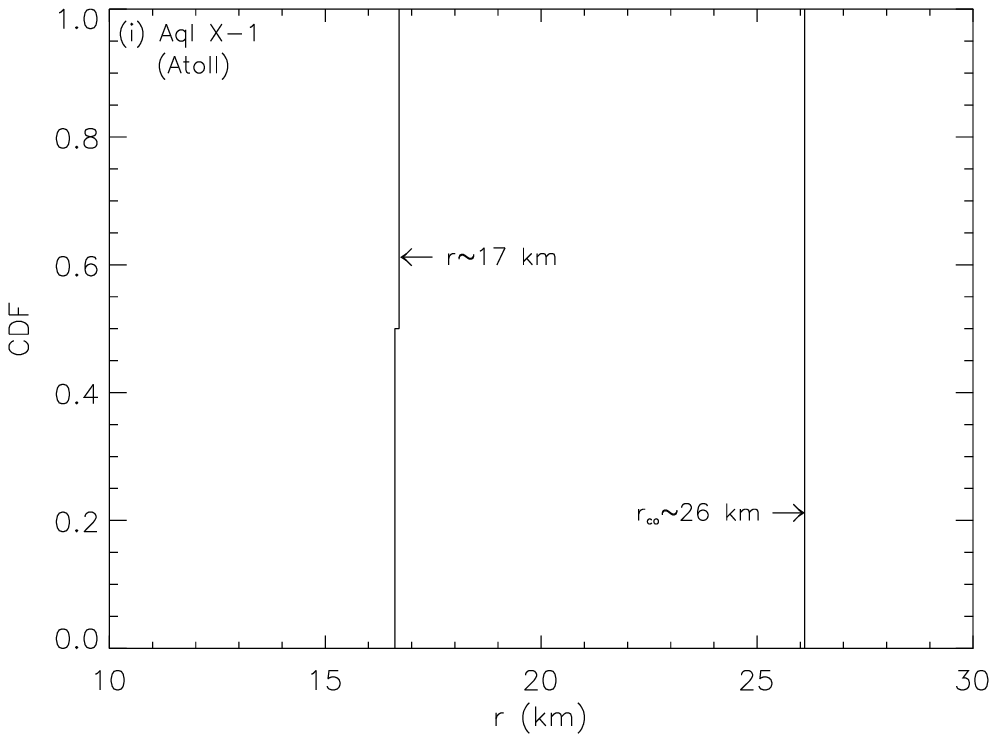}
\includegraphics[width=5.5cm]{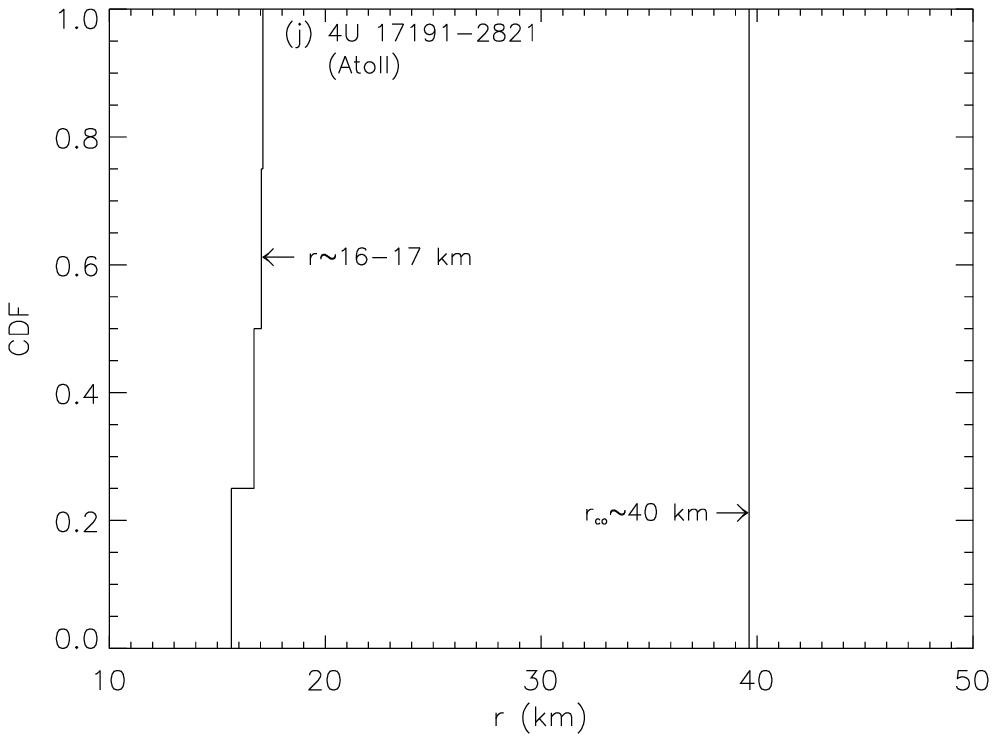}
\includegraphics[width=5.5cm]{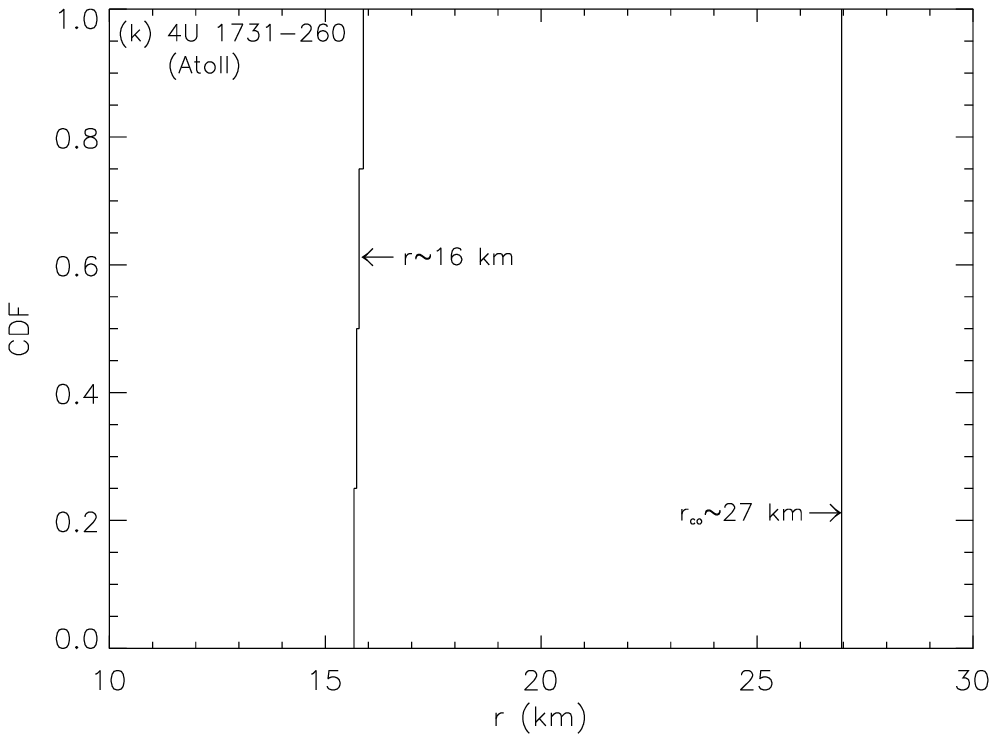}
\includegraphics[width=5.5cm]{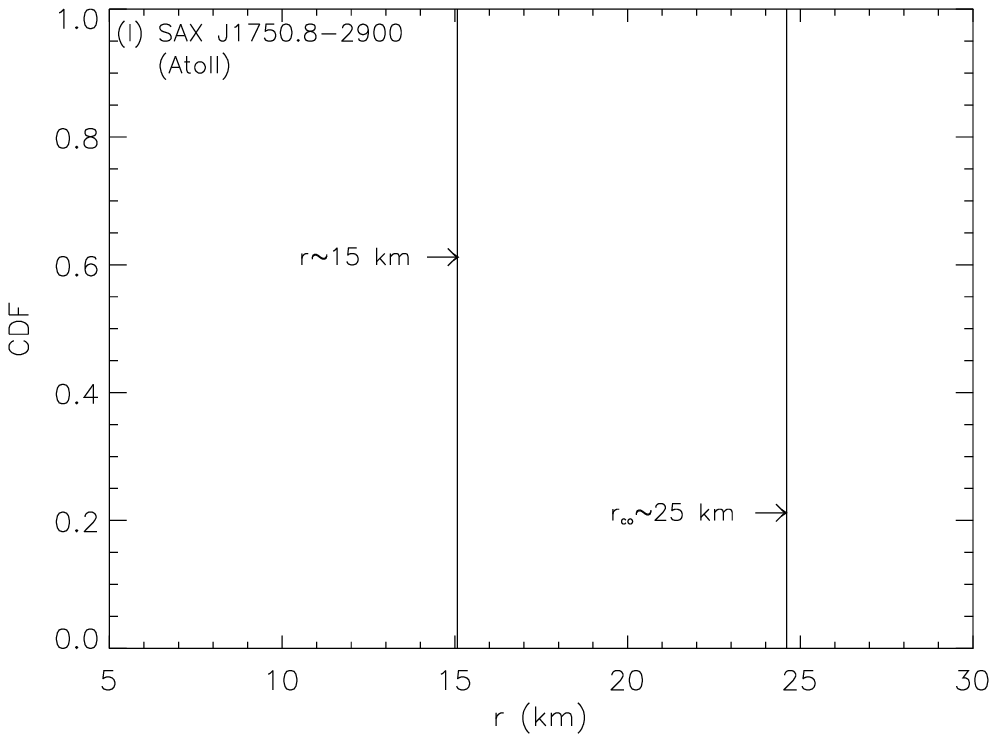}
\caption{ The CDF curve of the emission radius $r$ of twin kHz QPOs, as well as the position of the co-rotation radius $r_{\rm co}$, for
(a) SAX J1808.4-3658 ($r\sim21-24$\,km, $r_{\rm co}\sim31$\,km),
(b) XTE J1807.4-294 ($r\sim25-36$\,km, $r_{\rm co}\sim53$\,km),
(c) 4U 0614+09 ($r\sim16-30$\,km, $r_{\rm co}\sim31$\,km),
(d) 4U 1608-52 ($r\sim16-20$\,km, $r_{\rm co}\sim24$\,km),
(e) 4U 1636-53 ($r\sim15-20$\,km, $r_{\rm co}\sim25$\,km),
(f) 4U 1702-43 ($r\sim17$\,km, $r_{\rm co}\sim37$\,km),
(g) 4U 1728-34 ($r\sim16-25$\,km, $r_{\rm co}\sim34$\,km),
(h) 4U 1915-05 ($r\sim17-27$\,km, $r_{\rm co}\sim42$\,km),
(i) Aql X-1 ($r\sim17$\,km, $r_{\rm co}\sim26$\,km),
(j) IGR J17191-2821 ($r\sim16-17$\,km, $r_{\rm co}\sim40$\,km),
(k) KS 1731-260 ($r\sim16$\,km, $r_{\rm co}\sim27$\,km),
(l) SAX J1750.8-2900 ($r\sim15$\,km, $r_{\rm co}\sim25$\,km).
}
\label{r_r_co_source}
\end{figure*}

\section{Emission radius of twin kHz QPOs and co-rotation radius}

\subsection{Emission radius of twin kHz QPOs}

In this paper the both lower and upper kHz QPOs in one pair kHz QPOs are assumed to occur at the same radius, and the upper kHz QPO frequency $\nu_2$ is assumed as the Keplerian orbital frequency $\nu_{\rm K}$ (e.g. \citealt{Stella99a,Zhang04,van der Klis06}):
\begin{equation}
\nu_2=\nu_{\rm K}=\sqrt{\frac{GM}{4{\rm \pi}^2r^3}},
\label{nu_K}
\end{equation}
where $G$ is the gravitational constant, $M$ is the NS mass and $r$ is the Keplerian orbital radius, i.e. the emission radius of the kHz QPOs referring  to the center of NS.  By solving equation (\ref{nu_K}), the radius $r$ can be derived as:
\begin{equation}
\begin{aligned}
r&=(\frac{GM}{4{\rm \pi}^2})^{1/3}\nu_2^{-2/3}\\
&\approx19({\rm km})(\frac{M}{1.6\,{\rm M_\odot}})^{1/3}(\frac{\nu_2}{900\,{\rm Hz}})^{-2/3},
\end{aligned}
\label{r_K}
\end{equation}
where the mass  $1.6\,\rm M_\odot$ is the  average value of the millisecond pulsars \citep{Zhang11}, and frequency $900\,\rm Hz$  is the average frequency of the upper kHz QPOs (see \citealt{Wang14} for the details).
It is thought that kHz QPOs reflect the motion of matter in orbit at the inner accretion disk radius (or the magnetosphere-disk radius $r_{\rm m}$, see \citealt{van der Klis06}), i.e. $r\sim r_{\rm m}$. The magnetosphere-disk radius $r_{\rm m}$ is defined as the radius where the magnetic energy of NS becomes comparable to the kinetic energy of the accretion gas:\\
\begin{equation}
r_{\rm m}=\xi\,r_{\rm A},
\label{r_m}
\end{equation}
where $\xi$ is a constant factor of $\sim0.5$ in the thin accretion disk and $r_{\rm A}$ is the Alfv\'en radius \citep{Ghosh79,Shapiro83} ($\xi\sim1$ and $r_{\rm m}\sim r_{\rm A}$ in the spherical accretion, see \citealt{Bhattacharya91}). It is known that NS is in the spin-up state when $r_{\rm m}<r_{\rm co}$ while NS is in the spin-down state when $r_{\rm m}>r_{\rm co}$ (or $r_{\rm A}>r_{\rm co}$ for $\xi\sim1$, see \citealt{Bhattacharya91} for the details).

We infer the emission radii of the kHz QPOs in Table \ref{QPOs} by equation (\ref{r_K}) with the detected $\nu_2$ values and the assumed NS mass of $1.6\,{\rm M_\odot}$, where the NS mass of SAX J1808.4-3658 is adopted as $1.4\,{\rm M_\odot}$ by referring to its measured value  \citep{Elebert09a}.
The ranges of the inferred emission radii of kHz QPOs  of the 12 sources and their corresponding cumulative distribution function (CDF) curves are shown in Table \ref{QPOs} and Fig.\ref{r_r_co_source}, respectively.
We also show the CDF curve of the emission radii of all  kHz QPOs ($r\sim15-36$\,km) in Fig.\ref{r_r_co} (a), from which it can be seen that most emission radii cluster around the radius range of $\sim15-20$\,km ($\sim83$\% of the data), the rest of which mainly results from the source SAX J1808.4-3658 and XTE J1807.4-294 with the larger emission radii of $\sim21-24$\,km and $\sim25-36$\,km, respectively.

\subsection{Co-rotation radius}

\begin{figure}
\centering
\includegraphics[width=7.5cm]{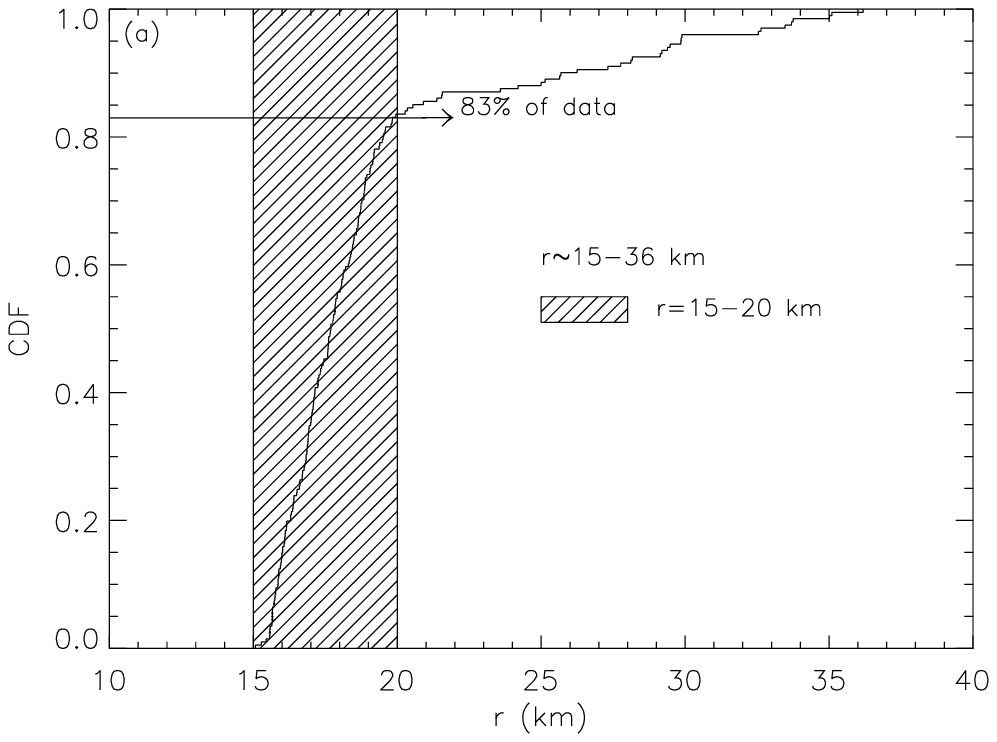}
\includegraphics[width=7.5cm]{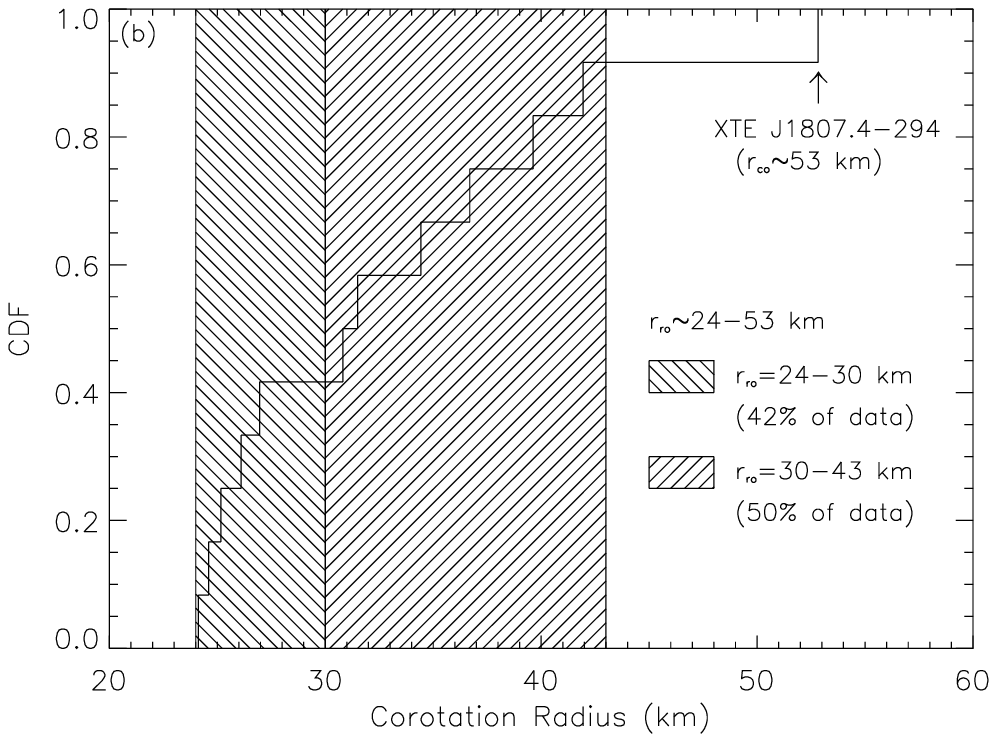}
\caption{(a) The CDF curve of the emission radius $r$ of all the twin kHz QPOs in Table \ref{QPOs} ($r\sim15-36$\,km). The shaded area shows the radius range of $15-20$\,km, which contains $\sim83\%$ of the data.
(b) The CDF curve of the co-rotation radius $r_{\rm co}$ of all the 12 sources in Table \ref{QPOs} ($r_{\rm co}\sim24-53$\,km). The shaded areas show the radius range of $24-30$\,km and $30-43$\,km respectively, which contains 42\% and 50\% of the data respectively. The arrow indicates the position of the co-rotation radius ($r_{\rm co}\sim53$\,km) of source XTE J1807.4-294.}
\label{r_r_co}
\end{figure}

The co-rotation radius $r_{\rm co}$ of the NS-LMXB \citep{Bhattacharya91} is the radial distance at  where the Keplerian orbital frequency   equals   the NS spin frequency (i.e. $\nu_{\rm K}=\nu_{\rm s}$). By setting  equation (\ref{nu_K}) to be  equal  to $\nu_{\rm s}$, $r_{\rm co}$ can be derived as:
\begin{equation}
\begin{aligned}
r_{\rm co}&=(\frac{GM}{4{\rm \pi}^2})^{1/3}\nu_{\rm s}^{-2/3}\\
&\approx32({\rm km})(\frac{M}{1.6\,{\rm M_\odot}})^{1/3}(\frac{\nu_{\rm s}}{400\,{\rm Hz}})^{-2/3},
\end{aligned}
\label{r_co}
\end{equation}

where  $\nu_{\rm s}$ is the NS spin frequency, and   $400\,\rm Hz$  is the average frequency of the detected spins of NS-LMXBs (see \citealt{Wang14} for the details).
We infer the co-rotation radii of the 12 sources in Table \ref{QPOs} by equation (\ref{r_co}) with the NS spin frequency $\nu_{\rm s}$ and the assumed NS mass of $1.6\,{\rm M_\odot}$.
The inferred $r_{\rm co}$ values of the 12 sources and their corresponding positions are shown in Table \ref{QPOs} and Fig.\ref{r_r_co_source}, respectively. We also show the CDF curve of all the 12 $r_{\rm co}$ values ($r_{\rm co}\sim24-53$\,km) in Fig.\ref{r_r_co} (b), from which it can be seen that 5 sources (42\% of the data) share the $r_{\rm co}$ range of $\sim24-30$\,km, 6 sources (50\% of the data) share the range of $\sim30-43$\,km, and the source XTE J1807.4-294 has the largest co-rotation radius of $\sim53$\,km. The average co-rotation radius of all listed sources is $\langle r_{\rm co}\rangle\sim32$\,km.

\subsection{Position parameter}

We introduce a ratio parameter $Y$ to study the relative position relation between the emission radius of the kHz QPOs and co-rotation radius quantitatively:

\begin{equation}
Y\equiv\frac{r}{r_{\rm co}}=(\frac{\nu_{\rm s}}{\nu_2})^{2/3},
\label{Y}
\end{equation}
which  depends on the frequencies of $\nu_2$ and $\nu_{\rm s}$.

\begin{figure}
\centering
\includegraphics[width=7.5cm]{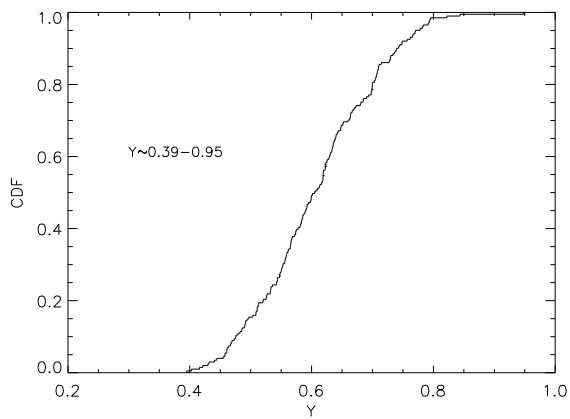}
\caption{
The CDF curve of all the position parameter $Y$ in Table \ref{QPOs} ($Y\sim0.39-0.95$).}
\label{delt_r_Y}
\end{figure}

For each pair of kHz QPOs in Table \ref{QPOs}, we calculate its corresponding position parameter $Y$ by equation (\ref{Y}) with $\nu_2$ and $\nu_{\rm s}$ values.
The ranges of the inferred $Y$ values of the 12 sources are shown in Table \ref{QPOs},  the CDF curve of which  is shown in Fig.\ref{delt_r_Y}.
The range of $Y$  is found to be $Y\sim0.39-0.95$, or all $Y$ values are less than unity.
In other words,  the emission radii of kHz QPOs of all the sources are smaller than their co-rotation radii.

\begin{figure}
\centering
\includegraphics[width=7.5cm]{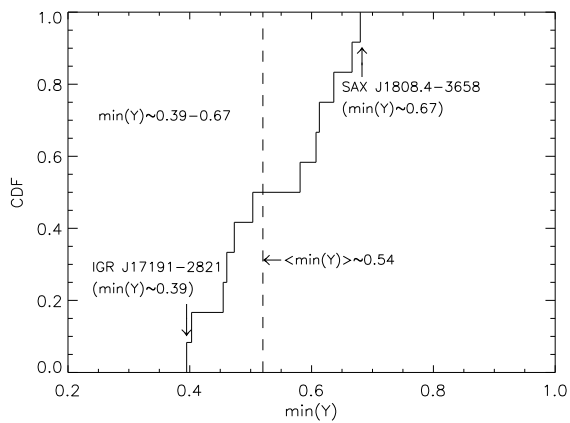}
\caption{
The CDF curve of the minima of the position parameter $Y$ for all the 12 sources (${\rm min}(Y)\sim0.39-0.67$ with the average value of $\langle {\rm min}(Y)\rangle\sim0.54$).
}
\label{min_r_r_co}
\end{figure}

Moreover, we also investigate  the innermost emission position of the kHz QPOs by analyzing the minimum of parameter $Y$ (min($Y$)). Fig.\ref{min_r_r_co} shows the CDF curve of the minima of parameter $Y$ of  12 sources, from which we notice
 that the minima of $Y$  lie in  the range of $\sim0.39-0.67$ with the average value of $\sim0.54$.

\section{Discussions and Conclusions}

Based on the data of   12 sources with the simultaneously  detected  twin kHz QPOs and NS spins,
 we investigate the relation between the emission radii of twin kHz QPOs and    co-rotation radii,
 and find that most of the emission positions of twin kHz QPOs cluster around   $\sim15-20$\,km, with
 the average co-rotation radius of  $\langle r_{\rm co}\rangle\sim32$\,km.
  The details of the conclusions and discussions  are summarized as follow:
\begin{figure}
\centering
\includegraphics[width=7.5cm]{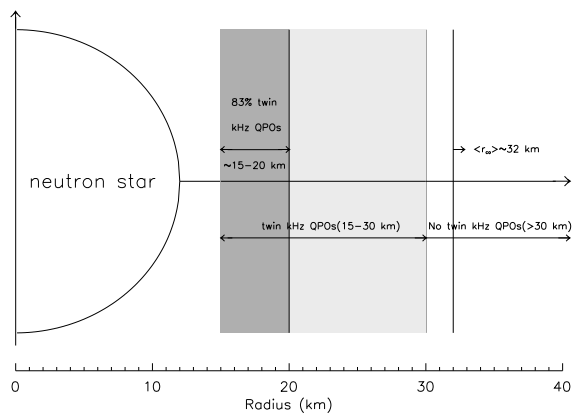}
\caption{The schematic diagram of the average co-rotation radius and the emission position of twin kHz QPOs. The average co-rotation radius of the 12 sources is $\langle r_{\rm co}\rangle\sim32$\,km, and the emission radii of twin kHz QPOs are in the range of $\sim15-30$\,km, and 83\% of which are in the range of $\sim15-20$\,km.}
\label{NS_QPO_r_co}
\end{figure}
\begin{enumerate}[(1)]
\item We analyze the ratios between the emission radius of the kHz QPOs and co-rotation radius ($Y\equiv\frac{r}{r_{\rm co}}<1$, see Table \ref{QPOs} and Fig.\ref{delt_r_Y} for the details), and find that all the  twin kHz QPOs are produced
     inside the co-rotation radii. This result indicates that the emission of twin kHz QPOs may be related to
     the spin-up process of accreting NS. Furthermore, $\sim83$\% of the inferred emission radii of twin kHz QPOs cluster around the NS
     surface at $r\sim15-20$\,km (see also Fig.\ref{r_r_co} (a)), which indicates that the twin kHz QPOs produce when the accreting matter collides  with the hard surface or environment with the local strong  magnetic field \citep{Zhang06}.
     In particular, 4U 1608-52 shows the emission radii of twin kHz QPOs to be $r\sim16-20$\,km (see Fig.\ref{r_r_co_source}d) while the corresponding lower kHz QPO frequencies are in range of 473-867\,Hz (as listed in Table \ref{QPOs}). For this wide range of lower kHz QPO frequencies, the X-ray spectrum changes significantly. For instance, \citet{Barret13} has shown that the comptonization parameters are related to the lower kHz QPOs for the source 4U 1608-52, and he also estimated the inner accretion disk radius in the range of $\sim15-25$\,km. Hence our proposed accretion disk structure is consistent with the result from the observed X-ray spectrum. It is noticed that parts of the sources in the sample are also observed the single kHz QPOs (e.g. \citealt{van Straaten00}), which show the larger frequency range (some ones lie in the frequency range of the upper kHz QPOs) compared with the twin kHz QPOs \citep{van der Klis00}. The larger frequency range indicates that the single kHz QPOs may have the larger range of the emission positions, i.e. inside or outside the co-rotation radius, hence implying that the occurrence condition of single kHz QPOs is not as rigorous as the twin ones.
     
\item The Keplerian frequency $\nu$ and the radius $r$ of the matter in the accretion disk has the following relation \citep{Zhang04,van der Klis06}:
    \begin{equation}
    \nu\sim r^{-3/2},
    \end{equation}
    which can be written into a variation form in the following: \\
    \begin{equation}
    \frac{\delta\nu}{\nu}\sim \frac{3}{2}\frac{\delta r}{r},
    \label{Keplerian}
    \end{equation}
    where $\delta r$ and $\delta\nu$ are set as $\delta r=r_{\rm co}-r$, $\delta\nu=\nu-\nu_{\rm s}$.
    The inferred minimum ranges of $\delta r$ for all the 12 sources are shown in Table \ref{QPOs},   min$(\delta r)\sim2$\,km. Substituting  $\langle r_{\rm co}\rangle\sim32$\,km and min$(\delta r)\sim2$\,km into equation (\ref{Keplerian}), one can obtain $\frac{\delta\nu}{\nu_{\rm s}}\sim\frac{3}{2}\frac{{\rm min}(\delta r)}{\langle r_{\rm co}\rangle}\sim10\%$, which indicates that  the twin kHz QPOs will not occur until the upper kHz QPO frequency is bigger than NS spin frequency by 10\%. We guess that  the emission of twin kHz QPOs may be related to the sufficient  velocity difference between the Keplerian motion of the accretion matter and  NS spin.
    Fig.\ref{NS_QPO_r_co} shows the schematic diagram of the emission position of twin kHz QPOs: As the accretion matter goes through the co-rotation radius, the  twin kHz QPOs will not emit until the distance difference satisfies  $\delta r\geq2$\,km.
     Therefore,  the twin kHz QPOs occur in-between the particular boundaries, NS surface and co-rotation radius,
      with the boundary layer of about several kilometers, which may represents the thickness of the transitional layer of accretion disk.

\item  As known, the less luminous source  SAX J1808.4-3658 shows the smaller frequency difference of twin kHz QPOs ($\sim1.5$ times smaller than other sources, see e.g. Wijnands et al.  2003),  and the  larger emission radii of its twin kHz QPOs are found  ($r\sim21-24$\,km).
    One explanation is that this  source may have the low accretion rate due to the low luminosity, which   causes  the accretion disk far away from the NS surface,  and the twin kHz QPOs emit at the farther positions.  In addition,  SAX J1808.4-3658 has been
    measured with the light NS mass ($<1.4\,{\rm M_\odot}$, see \citealt{Elebert09a}), which is smaller than the average value of the millisecond pulsars of 1.6 solar mass  \citep{Zhang11}, making its  Keplerian frequency systematically lower than those of other sources.
\end{enumerate}

\section*{Acknowledgments}

This work is supported by the National Basic Research Program of China (2012CB821800), the National Natural Science Foundation of China NSFC(11173034, 11173024, 11303047, 11565010), the Science and Technology Foundation of Guizhou Province (Grant No.J[2015]2113 and No.LH[2016]7226), the Doctoral Starting up Foundation of Guizhou Normal University 2014 and the Innovation Team Foundation of the Education Department of Guizhou Province under Grant Nos. [2014]35.

\bsp

\label{lastpage}

\end{document}